# Unveiling Linker-Born Electron Spin Centers in UiO-66-$NH_2$ MOF

Timur Biktagirov[1], Eugenio Otal[2], Leopold Trost[3], Patrick Dörflinger[4], Katsuya Teshima[1,5], Olga Trukhina[4], Daniel Klose[3], Wolf Gero Schmidt[2], Anastasiia Kultaeva[4]

*[1]Physics Department, Paderborn University, D-33098 Paderborn, Germany*

*[2]Institute for Aqua Regeneration (ARG), Shinshu University, 4-17-1 Wakasato, Nagano 380-8553, Japan*

*[3]Institute of Molecular Physical Science, ETH Zurich, 8093 Zurich, Switzerland*

*[4]Experimental Physics 6 and Würzburg-Dresden Cluster of Excellence ct.qmat, Julius-Maximilian University of Würzburg, 97074 Würzburg, Germany[5]Department of Materials Chemistry, Faculty of Engineering, Shinshu University, 4-17-1 Wakasato and Research Initiative for Supra-Materials (RISM) Interdisciplinary Cluster for Cutting Edge Research (ICCER), Shinshu University, Nagano 380-8553, Japan*

**Abstract:** Metal-organic frameworks (MOFs) offer a tunable platform for integrating functional spin centers into porous architectures with potential applications in catalysis and quantum sensing. Here, we identify a stable radical intrinsic to the aminoterephthalic acid linker in UiO-66-$NH_2$. Combining multi-frequency EPR and DFT, we unambiguously assign the long-misinterpreted EPR spectrum of UiO-66-$NH_2$ to this linker-centered NH• spin center. The radical can pre-exist in the linker precursor and remains stabilized within the MOF. At the same time, we explore the possibility of introducing this spin center post-synthetically. We examine its electronic structure and coupling to the nuclear spin environment in as-synthesized UiO-66-$NH_2$, underlining a new layer of quantum functionality in this widely used MOF. As a linker-based, chemically accessible species, this spin center could be introduced in other MOFs, offering a new platform for spin-defect engineering in catalysis, photophysics, and quantum sensing.

Metal-organic frameworks (MOFs), crystalline nanoporous materials composed of metal ions coordinated to organic linkers, have emerged as versatile platforms for a wide range of functional applications, including catalysis, sensing, gas storage, and separation technologies, owing to their high porosity, tunable chemistry, and exceptional structural stability [1]. Recently, significant attention has shifted towards exploring electron spin centers within MOFs due to their potential in diverse MOF applications. Electron spin centers can influence the electronic structure and chemical reactivity of MOFs, affecting charge separation and transfer processes that are pivotal in catalysis and other redox-driven transformations, as highlighted recently for photocatalytic MOF UiO-66 [2] and other frameworks [3–5]. Furthermore, these spin-bearing sites present opportunities for quantum technologies. They can enable quantum-enhanced sensing platforms capable of detecting guest molecules with high sensitivity spatial resolution [6–9]. Moreover, some spin centers in MOFs have been proposed as potential spin qubits due to their favorable coherence properties and controllable synthesis pathways [10–13].

Historically, functional electron spin centers in MOFs were predominantly associated with transition metal ions within their inorganic nodes. For example, $Cu^{2+}$ ions in Zn-doped HKUST-1 MOF have been shown to function as effective microwave-addressable quantum sensors [6, 14]. Similarly, metalloporphyrin units in MOFs have recently been explored as potential spin qubits [15–17]. However, recent interest has shifted to incorporating and leveraging organic radicals as spin centers bound to MOF linkers, due to long spin coherence times, straightforward synthesis, and tunable properties enabled by organic chemistry [7, 8, 18–20].

Here we present a particularly intriguing example of a stable organic spin center native to UiO-66-$NH_2$ [21], a prominent MOF material widely recognized for its compelling photocatalytic properties and exceptional chemical stability. We identify this spin center as a radical inherent to the MOF's linker 2-aminoterephthalic acid, TPA-$NH_2$ (also referred to as 2-amino-1,4-benzenedicarboxylic acid, BDC-$NH_2$).

A characteristic electron paramagnetic resonance (EPR) signal observed in UiO-66-$NH_2$ [22, 23] – and correlated with photocatalytic activity of the material – has conventionally been attributed to the metal clusters. Here, we combine multi-frequency continuous-wave (CW) and advanced pulsed EPR spectroscopy with density functional theory (DFT)-based spectral simulations to conclusively assign this EPR signature to a stable NH• radical spin center intrinsic to the TPA-$NH_2$ linker. Notably, we demonstrate – for the first time – that this radical is already present in free TPA-$NH_2$ and largely retains its spin properties upon incorporation into the MOF framework. At the same time, we also show that this spin center can be introduced post-synthetically. Thereby, our work corrects a longstanding misassignment that fundamentally alters the understanding of UiO-66-$NH_2$'s spin physics. Furthermore, since the TPA-$NH_2$ linker is inherent to a variety of MOFs beyond UiO-66-$NH_2$ (such as MIL and CAU families), our findings highlight a general new pathway for incorporating linker-based *intrinsic* spin centers in MOFs for promising applications, such as in microwave-addressable quantum sensing.

The UiO-66-$NH_2$ powder sample was synthesized according to the procedure described in the Methods section (see Supporting Information). The metal nodes of this MOF consist of zirconium oxide clusters, specifically $Zr_6O_4(OH)_4$ units coordinated by the carboxylate groups of the TPA-$NH_2$ linkers (Figure 1a). Most of the TPA-$NH_2$ molecules used in the synthesis are incorporated in UiO-66-$NH_2$ as linkers, while the remaining are washed out from the porous structure. This is supported by Figure 1b that presents the comparison of the steady-state photoluminescence (PL) emission spectra of the studied UiO-66-$NH_2$ sample, free TPA-$NH_2$ powder, and the sample of UiO-66 MOF that is formed of TPA without the $NH_2$ group.

The observed shift to lower wavelengths in the emission peak between UiO-66 and UiO-66-$NH_2$ indicates that the MOF sample incorporates TPA-$NH_2$ as linkers, with only a negligible amount of free TPA-$NH_2$ in the structure. This conclusion is further supported by the absence of the 620 nm PL peak associated with the pristine TPA-$NH_2$ linker. It is important to note that modifications to the MOF can lead to changes in the electronic structure, which is reflected in the shift of the PL peak maxima [24]. The overall PL intensity of the UiO-66-$NH_2$ sample is higher than that of UiO-66, with a photoluminescence quantum yield of 2.4% compared to 1.5%. This increase in PL intensity may indicate improved charge transfer between the ligands and metal clusters [24, 25]. However, due to the complex nature of PL, other factors such as crystallinity, particle size, and the presence of defects, among others, may also contribute to this behavior [25, 26].

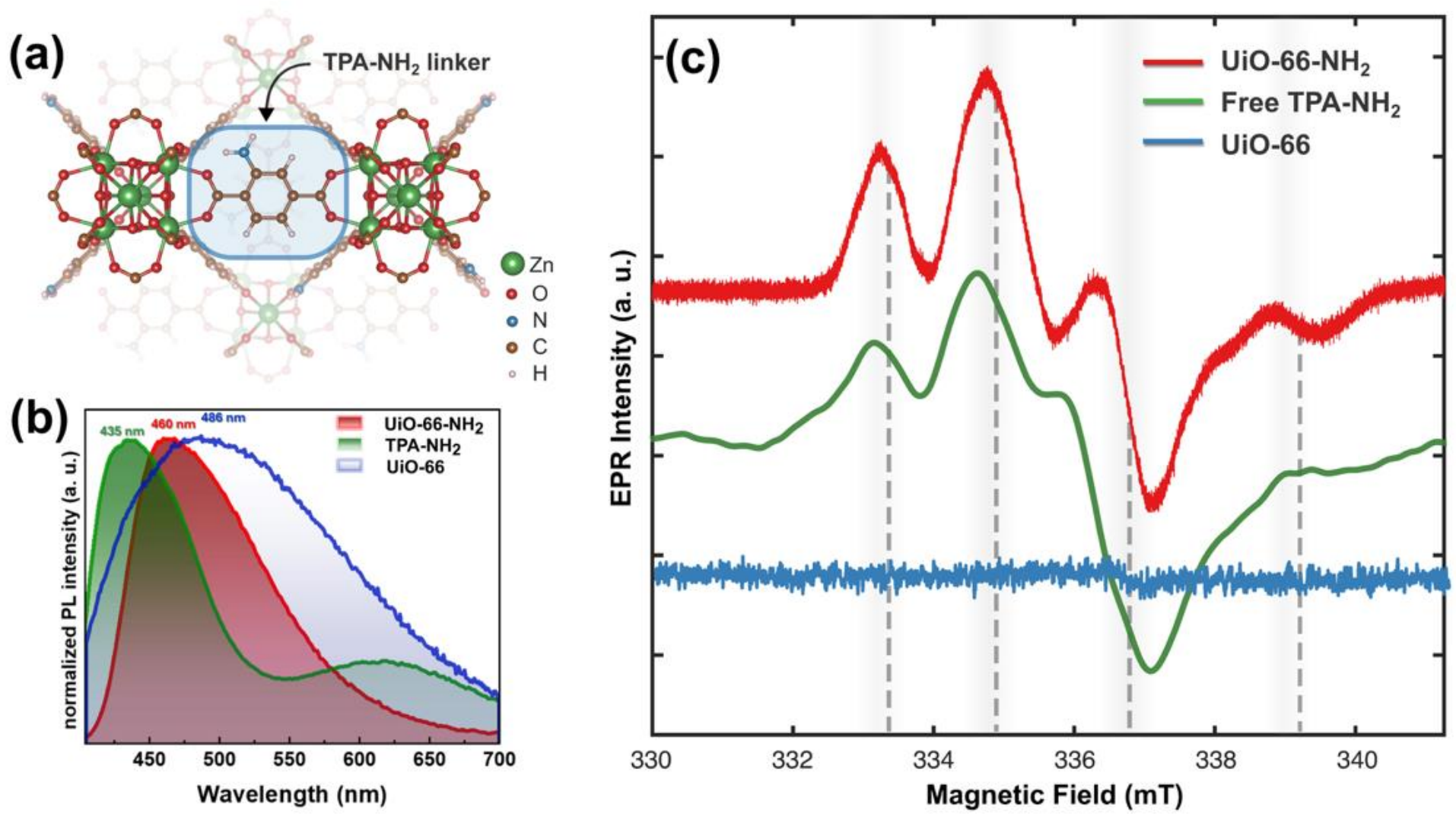

**Figure 1**. (a) Atomic structure of UiO-66-$NH_2$, with the arrow schematically indicating an electron spin center as a quantum sensor point on the linker. (b) Room temperature steady-state photoluminescence emission spectra of UiO-66-$NH_2$ MOF, free TPA-$NH_2$ powder and UiO-66 MOF. (c) Comparison of room-temperature X-band CW EPR spectra of UiO-66-$NH_2$ MOF, free TPA-$NH_2$, and UiO-66 MOF. The four-line splitting pattern associated with aminoterephthalate radicals is indicated with dotted lines.

In comparing the room-temperature CW EPR spectra of these samples (Figure 1c), we find that UiO-66-$NH_2$ and free TPA-$NH_2$ an identical four-line splitting pattern. At the same time, this EPR signature is absent in UiO-66, which features the same metal nodes as UiO-66-$NH_2$ but whose linkers are lacking the amino group. This unambiguously establishes that the spin center is associated with the TPA-$NH_2$ linker, not the MOF metal cluster, thereby resolving previous misassignment [22, 23]. The uniform saturation behavior upon increasing the microwave power observed for the four EPR lines (Supplementary Figure S3) supports that the lines originate from a single type of spin center, and not from a superposition of different species. Different types of centers would likely show varied saturation behaviors due to differences in their electron spin relaxation times [27]. Additionally, we found that the EPR spectral shape was not significantly affected by cooling to cryogenic temperatures (except for the appearance of additional signals of other spin centers near the free-electron *g*-factor, which is typical for UiO MOFs [2]; cf. Supplementary Figure S4), indicating that the radical is rigidly anchored within the MOF and does not exhibit detectable rotational mobility.

The resolved four-line splitting pattern can then be assigned to hyperfine (HF) interaction of an unpaired electron with a $^{14}N$ nucleus (nuclear spin $I$ = 1) and a single $^{1}H$ (nuclear spin $I$ = 1/2) nucleus, suggesting the presence of an NH• moiety that can form following hydrogen abstraction from the $NH_2$ group. The observation of the same EPR signal in free TPA-$NH_2$ (Figure 1c) suggests that the radical pre-exists in the linker precursor (as TPA-NH•) and may enter the MOF during synthesis. The radical concentration in as-purchased TPA-$NH_2$ powder estimated from the EPR spectrum intensity is about $5{\cdot}10^{-6}$ spins per molecule. In the MOF sample, the content of the radical remains of the same order of magnitude (approximately $3 \cdot 10^{-6}$ spins per linker). This suggests that once inside the framework, it remains spatially confined and electronically isolated by the rigid MOF lattice, which likely suppresses recombination pathways and kinetically stabilizes the radical [28].

While the formation of NH• in TPA-$NH_2$ has not been previously explored (to the best of our knowledge), other primary aromatic amines are generally known to undergo hydrogen abstraction from the $NH_2$ group, forming NH• radicals under ambient or mildly oxidative conditions. For example, aniline – the parent compound of TPA-$NH_2$, bearing the same aryl–$NH_2$ motif but lacking the carboxylic acid groups – readily forms NH• radicals upon photodissociation [29] or upon reactions with atmospheric free radicals such as OH• and Cl• [30–33]. Although the presence of electron-withdrawing carboxylic groups in TPA-$NH_2$ is expected increase the N-H bond strength [34], their effect should be moderate.

To illustrate this, we followed the modelling performed for aniline in Ref. [30] and computed the reaction barrier of OH•-mediated hydrogen abstraction from the $NH_2$ group on TPA-$NH_2$ (Supplementary Figure S5A). The results confirm that the barrier is only moderately higher compared to aniline. The calculated relative enthalpies of the transition state at 298 K are –0.04 eV for TPA-$NH_2$ and –0.14 eV for aniline. Similarly, the computed N–H bond dissociation enthalpy increases only modestly from 4.00 eV in aniline to 4.28 eV in TPA-$NH_2$, reflecting that NH• formation in TPA-$NH_2$ remains feasible. Notably, the alternative reaction pathway – OH addition to the aromatic ring that can also yield radical adducts [30] – involves considerably higher barriers and can be ruled out (Supplementary Table S2).

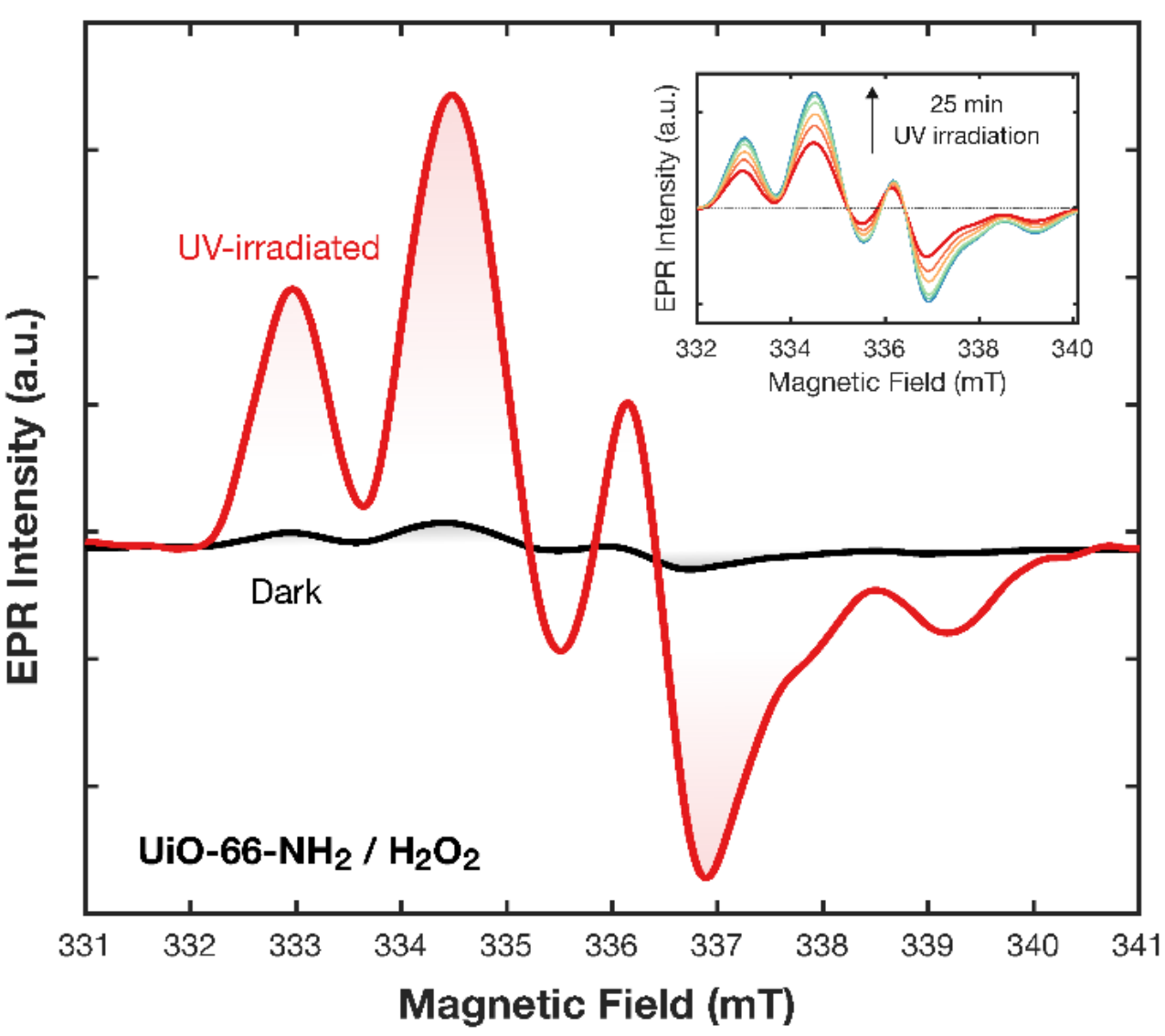

**Figure 2**. Room temperature EPR spectra of UiO-66-$NH_2$ in the presence of $H_2O_2$ (10 M) before (*black*) and after 10 min of UV irradiation (*red*), showing the effect of OH• radicals generated by $H_2O_2$ photolysis on the formation of the NH• spin centers. Continuous growth of the NH• EPR spectrum during the following 25 min of UV irradiation is shown in the inset.

Following these results, we explored the interaction of OH• with TPA-NH2 experimentally. In Supplementary Figures S5(A,B), OH• radical was generated by photolysis of $H_2O_2$, which led to the growth of the NH• EPR spectrum. Notably, a recent study [23] also observed the growth of the characteristic EPR spectrum in the UiO-66-$NH_2$ samples in the presence of $H_2O_2$, which was correlated to the formation of OOH• and/or OH• species. Following previous works, however, Ref. [23] tentatively assigned the EPR spectrum of NH• to a superposition of several distinct spin centers (namely, $Zr^{3+}$ ions, $Zr^{4+}$-bound superoxide anions, and amine-centered trapped holes). Regardless of this misattribution, the Ref. [23] vividly demonstrates that NH• can also form post-synthetically in UiO-66-$NH_2$. We confirm this by detecting NH• generation in the MOF by photolysis of $H_2O_2$, shown in Figure 2. These results not only demonstrate the possibility for controllable post-synthetic incorporation of the •NH spin center into the MOF but also underscore its potential role in catalysis and redox-driven transformations involving reactive oxygen species.

Next, to validate the proposed nature of the radical, we performed DFT modeling of NH• in UiO-66-$NH_2$ and calculated principal values of its $g$-factor and HF coupling constants (see the Methods section in Supporting Information). The DFT results reveal the appearance of localized radical-induced states in the electronic structure of the MOF (as illustrated by the density of states (DOS) plot shown in Figure 3a and the charge density plots in Figure 3b). Specifically, the NH•-bearing linker introduces a singly occupied π type electronic state in which the unpaired electron is in the $p$-orbital perpendicular to the plane of the linker. The highest fully occupied state of the radical is situated around 1 eV below the highest occupied states of the non-defective linkers (the formal valence band maximum) and features an electron residing in the $sp^2$ orbital ($\sigma_N$ type state). This aligns with the typical electronic structure of R-NH• radicals [35]. Notably, the presence of the half-filled mid-gap states localized on the linker may have implications for the photophysics of UiO-66-$NH_2$. Such spin defects can act as an electron or hole trap, influence charge separation dynamics [2], thereby affecting catalytic efficiency.

Another important result of our DFT modeling is that the electron-spin density of the spin center is well-confined within the single defective MOF linker, primarily on the NH• group with delocalization over the aromatic moiety (cf. Figure 3b). This explains why the EPR spectrum of the radical is largely unaffected by the framework and remains virtually the same as in free TPA-$NH_2$ powder.

Next, we used the results of DFT calculations to simulate the measured EPR spectrum according to the following spin Hamiltonian:

$$\hat{H} = \mu_B B \cdot g \cdot \hat{S} + \sum_i \hat{S} \cdot A_i \cdot \hat{I}_i \quad (1)$$

This allows establishing a direct connection between the proposed model of the spin center and the observed EPR signature. In Eq. (1), $\hat{S}$ is the electron spin operator (with the eigenvalue $S$ = 1/2), $g$ is $g$-tensor that reflects contributions from spin-orbit coupling to the free electron $g$-factor, $B$ is the magnetic field vector, $\mu_B$ is the Bohr magneton, $A_i$ is the tensor of HF interaction with $i$-th nucleus and $\hat{I}_i$ is the corresponding nuclear spin operator. The experimentally observed splitting between the four EPR lines (~2 mT) and their relative intensities suggest that the unpaired electron interacts with one $^{14}N$ and one $^{1}H$ nuclear spins with nearly equivalent effective HF coupling of about 60 MHz. Accordingly, the DFT results (listed in Supplementary Table S1) indicate that this splitting originates from the $^{14}N$ and $^{1}H$ nuclei of the NH• group, whose HF tensors feature the largest principal values of about 60

MHz. Subsequent simulation of the EPR spectrum using the entire set of DFT-calculated HF principal values shown in Figure 3d is in excellent agreement with the experiment, providing decisive evidence for the proposed nature of the radical.

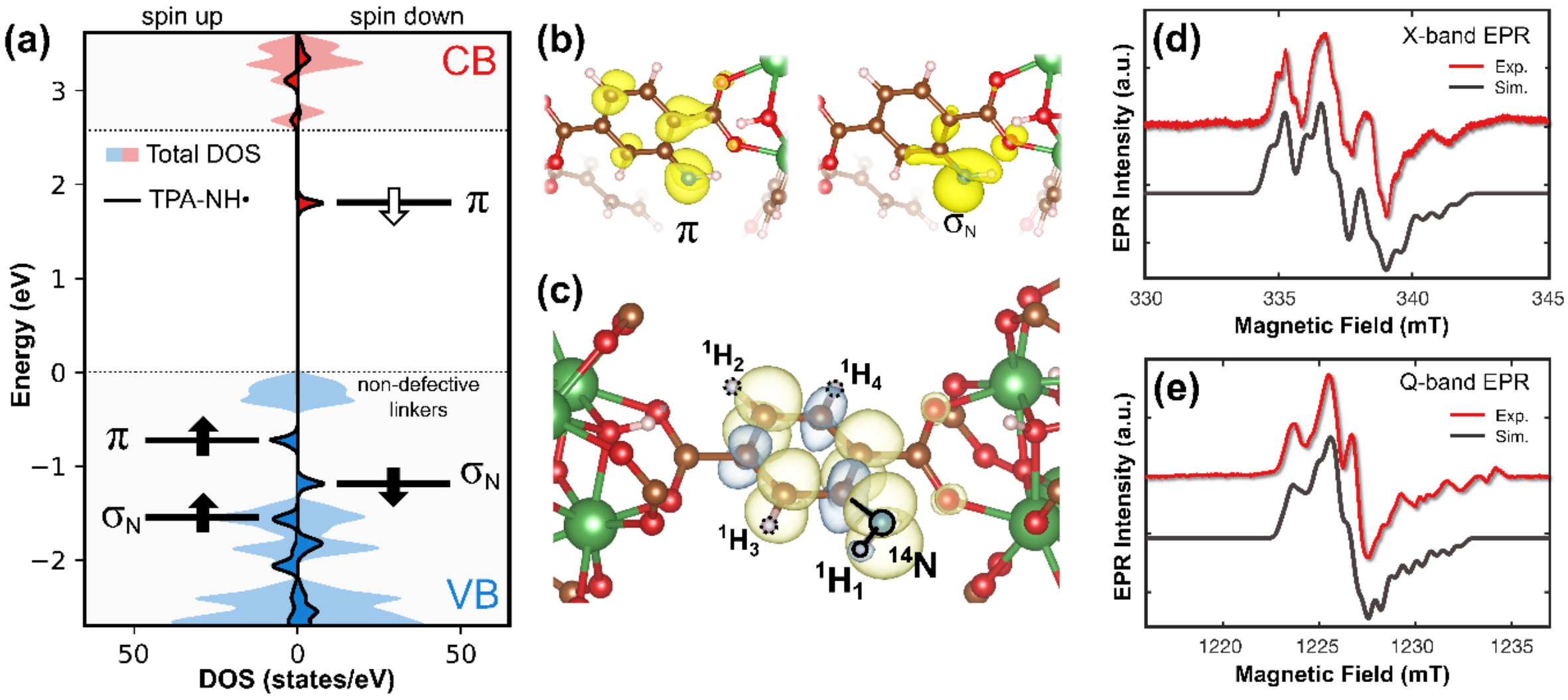

**Figure 3.** (a) DFT calculated total density of states (DOS; filled area; *blue*–occupied states, *red*–unoccupied states) and DOS projected onto the atoms of the radical-containing TPA-NH• linker in the UiO-66-$NH_2$ supercell (solid lines). Relevant Kohn-Sham states of the radical moiety are schematically shown with horizontal lines. Solid arrows depict occupied states, and an open arrow depicts an unoccupied state. (b) Charge density isosurfaces of the highest half-filled (π) and fully occupied ($\sigma_N$) orbitals of TPA-NH• in UiO-66-$NH_2$. (c) Electron spin density distribution of the NH• radical in UiO-66-$NH_2$. Yellow and blue lobes represent negative and positive isovalues, respectively; the isosurface absolute value is 0.002 e/Bohr$^3$. (d) Simulation of the X-band and (e) Q-band EPR spectra of UiO-66-$NH_2$ based on the DFT-calculated spin Hamiltonian parameters (listed in Table S1).

In the simulation, the Euler rotation angles of the HF tensors with respect to the *g*-tensor were optimized for a better fit with a maximum of 15° deviation from the DFT values (cf. Supplementary Table S1). Thus, not only the magnitude of HF couplings, but also the mutual orientation of their principal directions fit the experimental data. As for the HF splitting due to the distant $^1H$ nuclei of the aromatic group ($^1H_3$ and $^1H_4$ in Table S1 and Figure 3c) they are unresolved within the linewidth but contribute to the overall lineshape (cf. Supplementary Figure S2).

Lastly, we measured and simulated the EPR spectrum acquired in Q-band (at ~34 GHz) [36]. The higher magnetic field used in Q-band EPR provides enhanced resolution of the Zeeman splitting. Most importantly, a comparative analysis of Q-band and X-band spectra enables a clear distinction between *g*-factor anisotropy – which depends on the applied magnetic field – and HF splitting, which is field-independent (cf. Eq. (1)) [37]. In the simulation of the measured Q-band EPR spectrum of UiO-66-$NH_2$ shown in Figure 3e, we used the same set of spin Hamiltonian parameters established for the X-band spectrum, without any further adjustment. The good agreement between simulation and experiment provides definitive validation of our model. It is worth noting that all previous EPR measurements of UiO-66-$NH_2$ were restricted to X-band; our Q-band data thus provide the missing piece needed to resolve the longstanding ambiguity surrounding the interpretation of this EPR signature.

Having established the identity, electronic structure, and the formation pathways of the spin center, we next turn to X-band pulsed EPR measurements at 50 K to explore how NH• interacts with its surrounding nuclear spin environment. The echo-detected field-sweep (EDFS) spectrum of UiO-66-$NH_2$, obtained using the Hahn-echo pulse sequence is shown in Supplementary Figure S6. Its close similarity with the first integral form of the CW EPR spectrum indicates that the same spin centers are probed by both techniques and that all relevant transitions are effectively excited by the pulsed experiment.

The Hahn-echo decay as a function of the inter-pulse delay $\tau$ shown in Figure 4a can be accurately described by a single stretched exponential of the form $I(2\tau) = exp\,(-[2\tau/T_m]^{\beta})$. This behavior confirms that the signal arises from a single spin species rather than a superposition of several centers with differing relaxation properties. The inversion recovery experiment (Figure S7) likewise supports this conclusion by yielding a monoexponential recovery of magnetization. The extracted phase memory time, $T_m$ = 1.129 µs, characterizes how long the spin coherence of NH• is preserved in as-synthesized UiO-66-$NH_2$ before dephasing due to interactions with surrounding nuclear spins and nearby electron spins. The stretching parameter, $\beta$ = 0.91, reflects a moderate distribution of coherence times – indicative of an inhomogeneous spin bath environment [38–40].

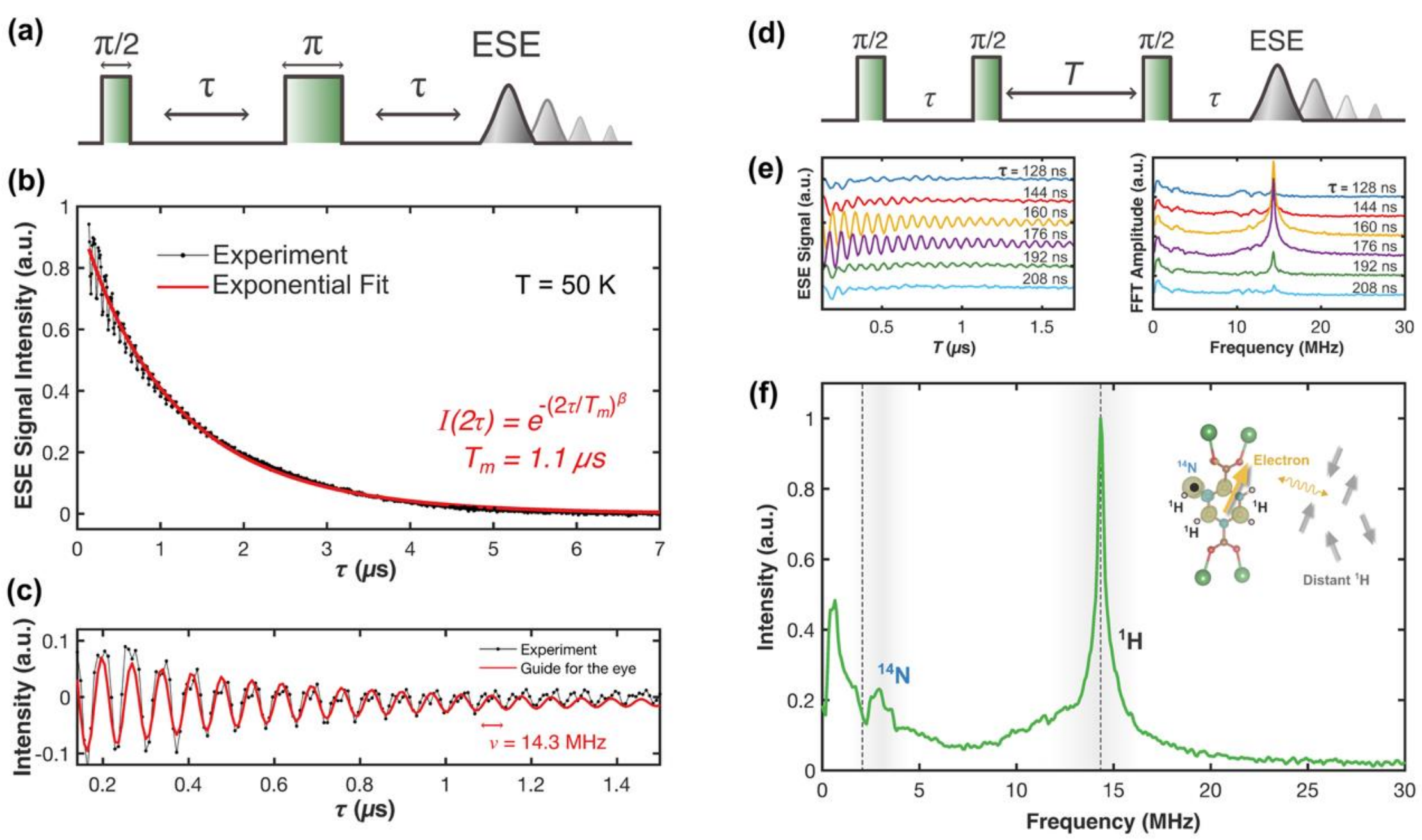

**Figure 4.** (a) Hahn-echo pulse sequence and (b) Hahn-echo decay of the NH• spin center measured at 50 K as a function of the time delay $\tau$ between the microwave pulses. A stretched exponential fit shown with a red trace. (c) Modulations of the decay curve (with a frequency of 14.3 MHz) after subtracting the baseline. (d) 3-pulse ESEEM pulse sequence. (e) Time traces measured at 6 different $\tau$-values and the corresponding Fourier-transformed magnitude spectra. (d) ESEEM sum spectrum; hydrogen transitions are dominant and found at the corresponding Larmor frequency (14.3 MHz, labeled as $^1$H). A peak assigned to $^{14}$N double quantum transitions showing the HF coupling is marked as $^{14}$N, and the corresponding Larmor frequency (2 x 1.03 MHz) is indicated with a dotted line.

The observe microsecond-scale $T_m$ at 50 K is in line with values reported for spin centers in nuclear-spin-rich environments, such as in solvent-filled MOFs or in molecular systems surrounded by spin-bearing solvent molecules [12, 41, 42]. Insights into the nuclear spin environment beyond the nearest nuclei – which are already probed via CW EPR and DFT – can be obtained from the analysis of the oscillatory behavior of the Hahn-echo decay seen in Figure 4a. This effect, known as electron spin echo envelope modulation, ESEEM [43], is more clearly visible after subtraction of the stretched exponential baseline (Figure 4b). The modulation depth and frequency provide information about the number, type, and distance of nuclei coupled to the unpaired electron [44,45]. The dominant modulation frequency observed in Figure 4b is close to the Larmor frequency of $^{1}$H ($v_L^H \approx 14.3$ MHz at the experimental magnetic field), suggesting coupling to both linker protons and more distant hydrogen-containing solvent molecules contained in the pores.

To further resolve the nature of these nuclear interactions, we performed 3-pulse ESEEM spectroscopy (Figure 4d-f). In this experiment, the envelope decay is defined predominantly by nuclear relaxation, which is substantially longer compared to the electron $T_m$ time in the Hahn-echo measurements, thus enabling a significantly better spectral resolution [45]. Time traces acquired at multiple values to suppress "blind spot" artifacts were Fourier-transformed (Figure 4e) and averaged, yielding the final ESEEM spectrum (Figure 4f). The most prominent feature of the ESEEM spectrum is a signal centered around $v_L^H$, associated with hyperfine-coupled $^{1}$H nuclear spins. Each nuclear spin contributes to the ESEEM spectrum at both sides of the corresponding Larmor frequency with the shift defined by its individual HF coupling tensor [45]. Since the magnitude of the dipolar contributions to the HF interaction depends on the distance between the electron and nuclear spins, broad shoulders around $v_L^H$ are caused primarily by the $^{1}$H atoms located in the MOF structure near the NH• group [6]. At the same time, the sharp peak at $v_L^H$ signifies that the $^{1}$H nuclear spin ensemble coupled to the electron spin of NH• is dominated by the distribution of distant nuclei, which we associate with solvent molecules distributed in the pore system of UIO-66-$NH_2$.

Additionally, a much weaker feature observed in Figure 4f near twice the Larmor frequency of $^{14}$N was assigned to double quantum transitions of hyperfine-coupled $^{14}$N. This signal is consistent with similar 3-pulse ESEEM features reported in related systems [46, 47], and provides further supporting evidence for the involvement of distant nitrogen nuclei in the spin environment of the radical.

Taken together, these data demonstrate that the NH• radical in the as-synthesized MOF is embedded in a dense nuclear spin environment comprising both framework and solvent $^{1}$H and $^{14}$N nuclei, which limits the $T_m$ timescale via spin diffusion. We emphasize that our aim here was not to optimize the center's coherence but to establish a realistic baseline for spin dynamics in as-synthesized UiO-66-$NH_2$. However, further optimization, including the exchange or the removal of solvent from pores through MOF activation, may lead to significantly enhanced coherence times. As demonstrated in previous studies, coherence times of molecular spin centers can be extended significantly when using spin-free solvents [41].

In summary, we unambiguously identified a native radical defect in UiO-66-$NH_2$ and established its localization on the amino-functionalized linker. Our results resolve a long-standing misassignment of the EPR signature to metal-cluster-bound species, clarifying the fundamental nature of intrinsic spin centers in this archetypal MOF. This chemically accessible, framework-stabilized spin center with a well-defined spin signature endows UiO-66-$NH_2$ – a MOF widely used in catalysis, sensing, and related applications – with a new layer of quantum functionality. The NH• radical is present already in free TPA-$NH_2$ and largely retains its properties upon incorporation into the MOF. This suggests that NH• formation is governed by the inherent chemical reactivity of the linker, and it may therefore be introduced into other MOF topologies constructed from TPA-$NH_2$. These findings open a pathway for the rational design of functional organic spin centers in MOFs – not only for emerging quantum sensing applications, where spatial confinement and coherence are essential, but also for established MOF domains such as catalysis, redox chemistry, and photophysics, where linker-localized radicals may play an active role in charge transfer and reactivity.

**SUPPORTING INFORMATION**

Full experimental and computational details and additional spectroscopic data (PDF)

**ACKNOWLEDGEMENTS**

The authors thank Professor Vladimir Dyakonov for fruitful discussion. DK gratefully knowledges financial support from ETH research grant ETH-35221. TB and WGS acknowledge the Paderborn Center for Parallel Computing (PC2) for the provided computational resources. AK acknowledges financial support from the Würzburg-Dresden Cluster of Excellence on Complexity and Topology in Quantum Matter ct.qmat (EXC 2147, DFG project ID 390858490). P.D. acknowledges the German Research Foundation (DFG) program SPP2196 under DY18/14-2 (Project number 424101351).

**SUPPORTING INFORMATION for**

# Unveiling Linker-Born Electron Spin Centers in UiO-66-$NH_2$ MOF

Timur Biktagirov[1], Eugenio Otal[2], Leopold Trost[3], Patrick Dörflinger[4], Olga Trukhina[4], Daniel Klose[3], Wolf Gero Schmidt[2], Anastasiia Kultaeva[4]

## FULL EXPERIMENTAL AND COMPUTATIONAL DETAILS

### Sample preparation

UiO-66-$NH_2$ was synthesized using a modified procedure previously reported by Katz et al. [1]. In a 500 ml Schott Duran flask, 300 ml of DMF and 20 ml of concentrated HCl were mixed under a fume hood. To this solution, 2.51 g of $ZrCl_4$ was added in small portions, also under the fume hood. The solid dissolved quickly, and then 2.68 g of TPA-$NH_2$ was added to the resultant solution. To facilitate the dissolution of TPA-$NH_2$, the mixture was subjected to an ultrasonic bath for 30 minutes. After complete dissolution, the screw-cap bottles were sealed and maintained at 120°C for 12 hours. The solids were isolated by centrifugation and washed twice with DMF and $CH_2Cl_2$.

### Continuous wave electron paramagnetic resonance (CW EPR) spectroscopy

The X-band CW EPR spectra (9.42 GHz) were measured on a commercial Magnettech spectrometer equipped with an Oxford ESR 900 He flow cryostat. A microwave power of 1 mW was chosen for optimal signal-to-noise ratio of the main EPR spectra without saturation effects. Experiments on the saturation behavior and temperature dependence of the EPR spectrum have also been conducted (see Figure S2 and Figure S3). A modulation amplitude of the external magnetic field of 2 Gauss was used to resolve the hyperfine EPR lines. To determine the spin concentration, experiments were first conducted to saturate the EPR spectra with microwave power, and the amplitude of the magnetic field was optimized to eliminate any possible spectral line shifts during the measurements. Spin counting was performed at room temperature. Radical concentration was calculated by the standard spin counting procedure via double integration of the CW spectra and comparison with a calibrant (DPPH and Ultramarine samples).

EPR spectra in Q-band were recorded on an Elexsys E580 EPR spectrometer equipped with a cryogen-free variable temperature cryostat (Cryogenic Ltd., London, UK) and a home-built dedicated CW EPR resonator for oversized sample tubes [2]. Data were acquired using 1001 points over a width of 50 mT, 0.005 mW microwave power to ensure non-saturating conditions, a lock-in conversion time 160.0 ms and a time constant 40.96 ms, a magnetic field modulation amplitude 0.05 mT and a modulation frequency 100 kHz, 240 scans were accumulated.

### Pulsed EPR spectroscopy

Pulsed X-band EPR experiments were performed on an Elexsys E680 EPR spectrometer (Bruker) with a dielectric ring resonator (MD-5, Bruker), a 1 kW TWT amplifier (Applied Systems Engineering Inc., Texas, USA) and a Helium flow EPR cryostat (CF935, Oxford Instruments) to maintain the temperature at 50 K.

The echo-detected EPR spectrum was measured using the Hahn echo sequence, $\pi/2$ - $\tau$ - $\pi$ - $\tau$ - echo, with an interpulse delay $\tau$ of 200 ns, 16/32 ns rectangular pulses for $\pi/2$ and $\pi$, respectively. The echo was integrated with a boxcar integration window of 200 ns. Data were acquired by stepping the magnetic field over 50 mT in 501 points, using 100 shots per points with a shot repetition time of 8 ms and 10 scans were accumulated. 2pESEEM data were acquired by using the same sequence at the magnetic field corresponding to the maximum of the absorption spectrum and stepping the interpulse delay τ in 1024 steps of 8 ns from 140 ns using a two-step phase cycle (+x - -x) on the π/2-pulse with a boxcar echo integration window of 32 ns, 10 shots per points and a shot repetition time of 2.6 ms, while 240 scans were accumulated.

3pESEEM data were acquired using the pulse sequence, π/2 - τ - π/2 - T - π/2 - τ - echo. Data were acquired for six different τ-values by using the same sequence at the magnetic field corresponding to the maximum of the absorption spectrum and stepping the interpulse delay T in 512 steps of 8 ns from 128 ns using a four-step phase cycle with a boxcar echo integration window of 32 ns, 10 shots per points and a shot repetition time of 3 ms, while 420 scans were accumulated.

The longitudinal relaxation time, $T_1$, was measured using the inversion recovery sequence, π - t - π/2 τ - π - τ - echo, with an interpulse delay τ of 200 ns, 16/32 ns rectangular pulses for π/2 and π, respectively, with a two-step phase cycle (+x - -x) on the π/2-pulse to cancel out receiver offsets. The echo was integrated with a boxcar integration window of 32 ns. Data were acquired by stepping t from 1.2 µs in steps of 20 µs for 2048 points, using 4 shots per point with a shot repetition time of 53 ms and 9 scans were accumulated.

**Computational methods**

DFT calculations of spin Hamiltonian parameters were performed with the ORCA (v. 6.0) program package [3] using the PBE exchange-correlation functional [4] and the def2-TZVP basis set [5]. The cluster model used in the calculations was cut from the crystal structure of UiO-66. It consisted of one linker and two SBUs with the remaining (unsaturated) coordination sites of the SBUs terminated by $HCOO^-$ anions. During geometry optimization, the coordinates of the carbon atoms of each $HCOO^-$ were fixed. To address the electronic structure of the radical-containing $NH_2$-UiO-66, a 126-atom MOF supercell with one spin center was modeled with the Quantum ESPRESSO package [6, 7] under periodic boundary conditions. The calculations used the hybrid HSE06 functional [8, 9], projector augmented wave pseudopotentials [10], 50 Ry kinetic energy cutoff, and *Γ*-point Brillouin zone sampling.

The simulations of the powder EPR spectra were carried out using the Matlab toolbox EasySpin [11]. The simulations were based on the spin Hamiltonian defined in Eq. (1) that included the anisotropic Zeeman interaction, and the anisotropic HF interactions between the electron spin and the $^{14}$N nuclear spin, as well as to the four $^{1}$H nuclear spins of the radical-containing linker of $NH_2$-UiO-66. The DFT-calculated spin Hamiltonian parameters were used as a starting point for EPR spectra simulations. During the simulations, the principal values of the *g*-tensor and the Euler rotation angles of the HF tensors were adjusted to obtain the best fit. The fitted spin Hamiltonian parameters are provided in Table S1.

**Photoluminescence**

For the steady-state photoluminescence (SSPL) measurements an Edinburgh Instrument FLS 980 was used. The powder sample was measured in a reflection geometry and the emission of the sample was collimated. In front of the monochromator and the photomultiplier tube a 400 nm longpass filter was positioned, which blocks wavelengths below 400 nm. To measure the steady-state PL spectra, the samples were excited with a 375 nm laser having a spot size of 100 µm in diameter, a fluence of approximately 88 nJ/cm², and a repetition rate of 50 ns. For the PLQY measurements, an integrating sphere in an Edinburgh Instruments FLS 980 was used. The sample was excited with a 406 nm CW laser diode with 31.9 mW $cm^{-2}$. The spot size was 0.03 $cm^2$ [12].

**pXRD**

X-ray diffraction pattern was obtained using a Rigaku MiniFlex diffractometer with Cu-Kα radiation and a step size of 0.0100 degrees. No significant broadening of the peaks was found, indicating that the crystallinity of the X-ray diffraction pattern.

**ADDITIONAL DATA**

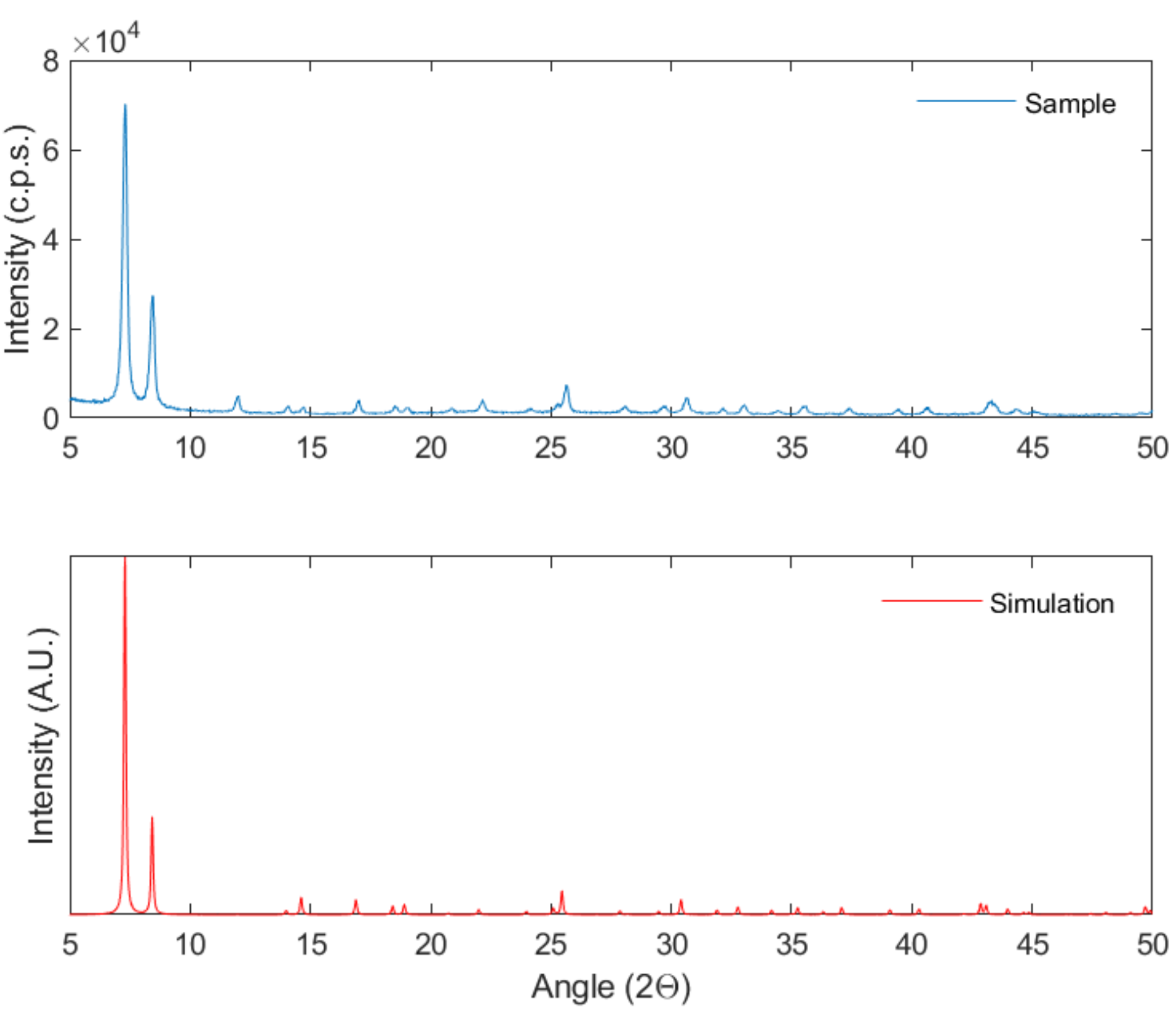

**Figure *S*1.** X-Ray diffraction patterns for UiO-66-$NH_2$ powder sample.

**Table S1**. DFT calculated spin Hamiltonian parameters of TPA-NH• in UIO-66-$NH_2$ along with the parameters used to simulate the experimental X-band and Q-band EPR spectra in Fig. 3d,e (same parameters used for both X- and Q-band). Simulation parameters are only explicitly given in the table if they were varied to fit the experimental spectra - otherwise the DFT values were used and kept fixed in the simulations.

| | | Principal Values* | Euler Rotation Angles to the *g*-tensor** |
|---|---|---|---|
| *g*-tensor | DFT | 2.0026, 2.0032, 2.0053 | |
| | Sim | 2.0027, 2.0062, 2.0092 | |
| $^{14}N$ HFI | DFT | 59.59, -14.17, -13.39 | -146.7, 10.2, 121.3 |
| | Sim | *fixed* | -143.2, 9.2, 123.95 |
| $^{1}H_1$ HFI | DFT | -35.36, -58.13, 2.27 | -108.1, 33.3, 81.0 |
| | Sim | *fixed* | -108.5, 33.3, 80.4 |
| $^{1}H_2$ HFI | DFT | -17.11, -27.53, -6.20 | 93.1, 37.8, -99.3 |
| | Sim | *fixed* | 91.1, 39.9, -100.0 |
| $^{1}H_3$ HFI | DFT | -16.10, -20.84, -5.31 | -78.5, 32.8, 78.3 |
| | Sim | *fixed* | -72.7, 41.5, 69.7 |
| $^{1}H_4$ HFI | DFT | 1.30, 5.64, 4.21 | -108.0, 18.4, 101.9 |
| | Sim | *fixed* | *fixed* |

*Principal values of the hyperfine tensors are in MHz
**Euler rotation angles are in degrees

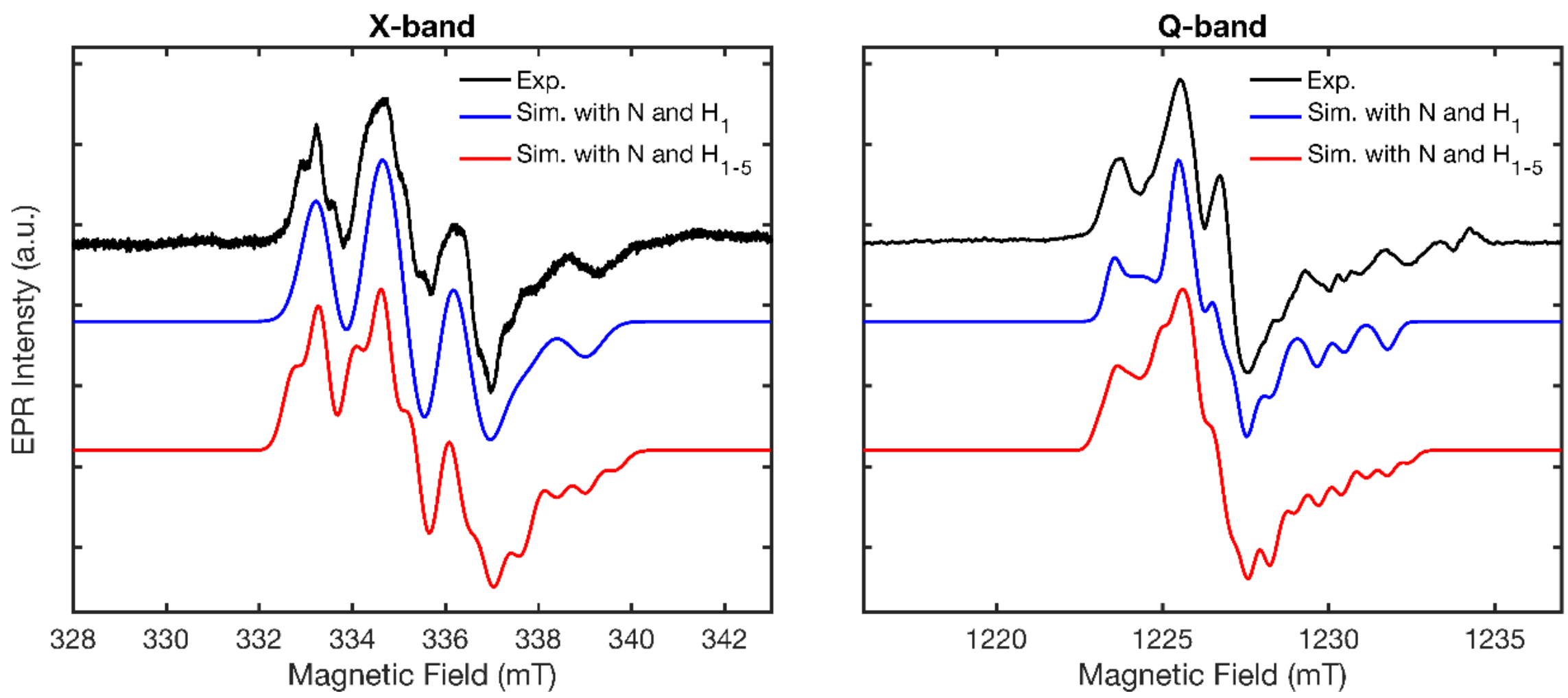

**Figure S2**. X-band (left) and Q-band (right) simulations of CW EPR spectra with the parameters listed in Table S1, including the contributions of only the two nuclei of the NH• group ($^{14}N$ and $^{1}H_1$; *blue line*) and all nuclei of the TPA-NH• linker (*red line*).

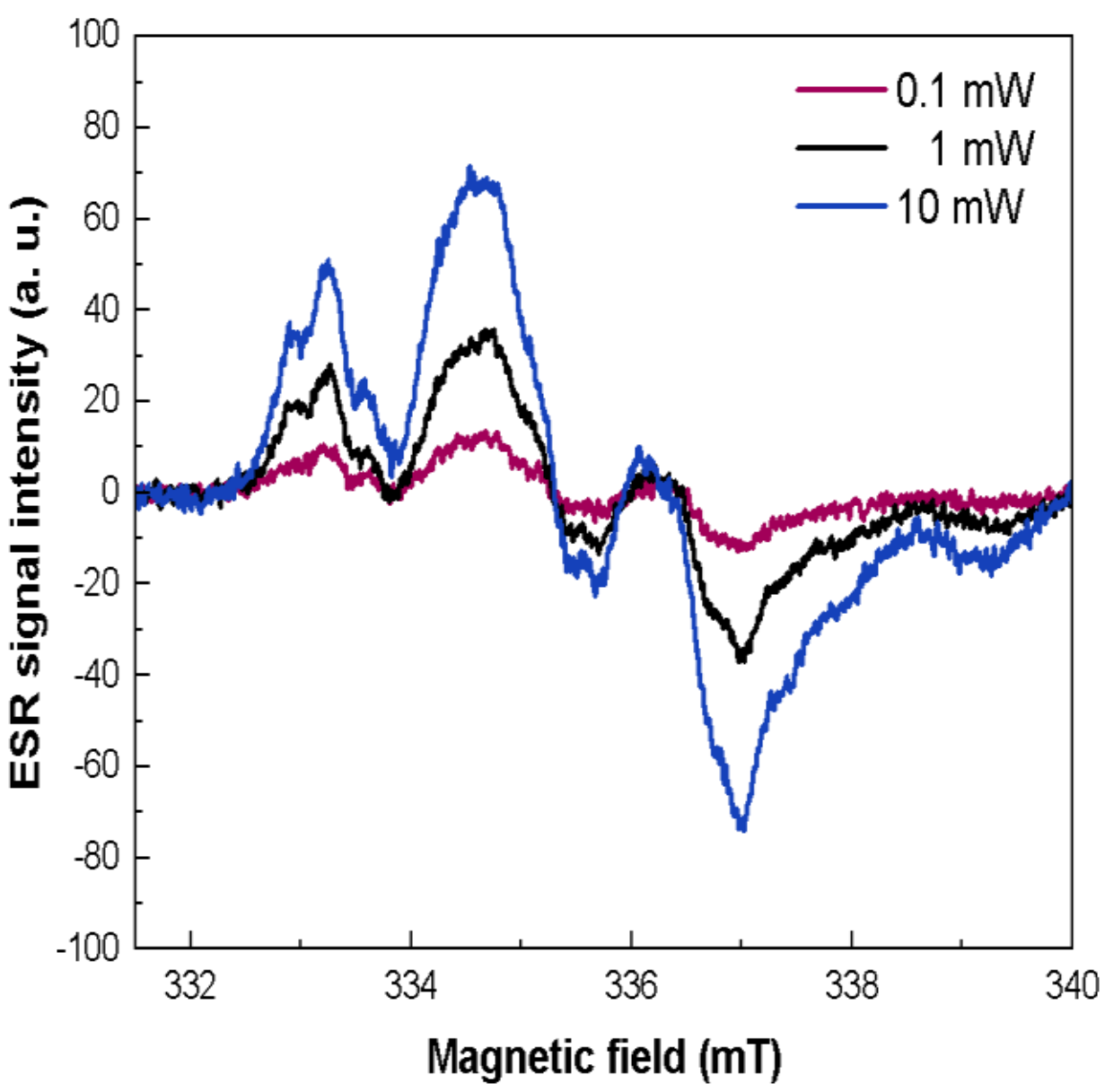

**Figure S3**. Saturation behavior of the CW EPR spectrum of UiO-66-$NH_2$. The microwave power saturation experiment demonstrates a simultaneous increase in the EPR lines, which confirms the absence of superposition of lines with different nature (relaxation times and coordination geometry). The measurements were performed at room temperature.

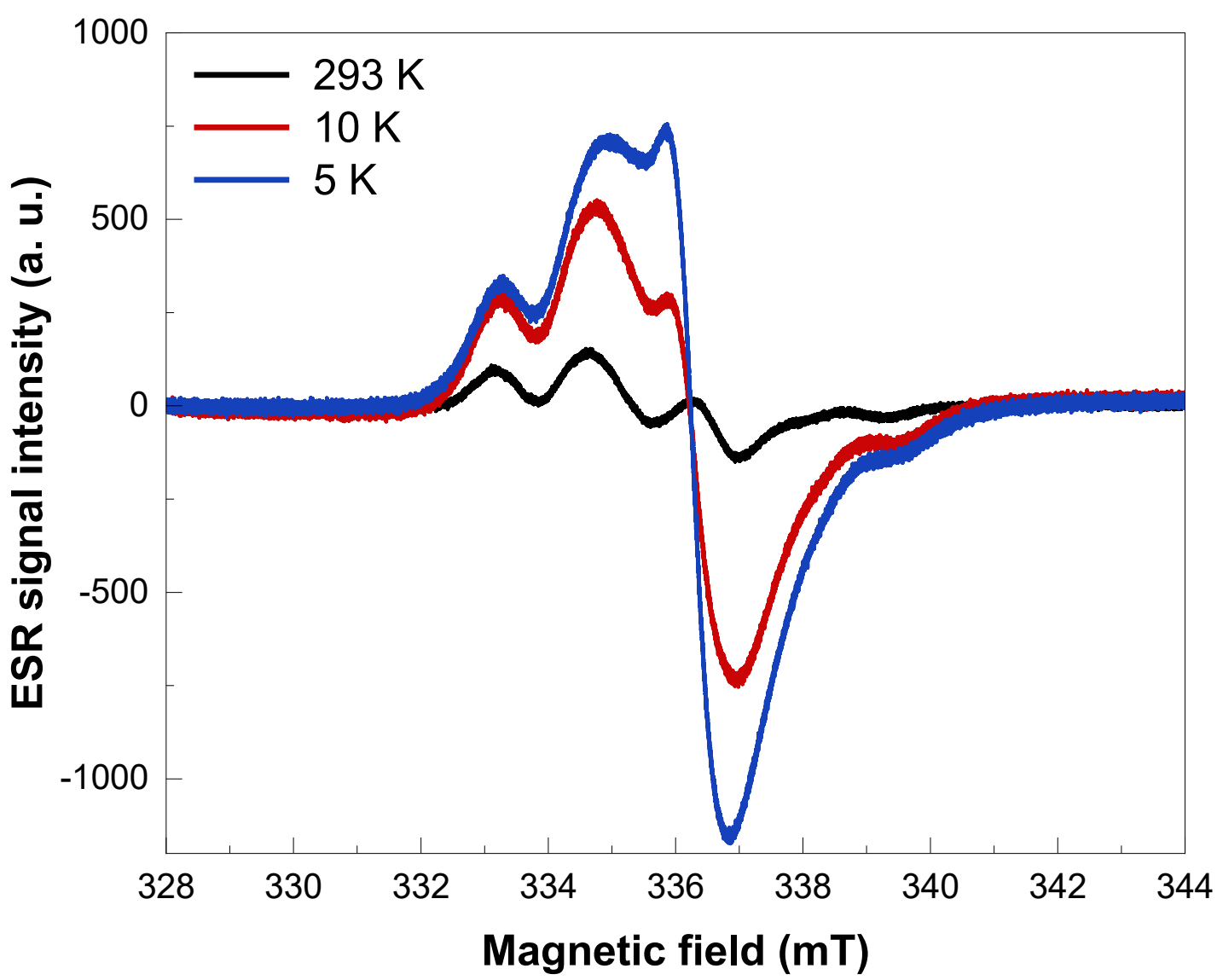

**Figure S4**. Temperature dependence of the CW EPR spectrum of UiO-66-$NH_2$, measured at the same experimental conditions. A simultaneous increase in the intensities of the EPR peaks is observed with decreasing temperature. At 5 K in the central part of the spectrum, a single line, whose g-factor is close to 2.0023, also begins to appear.

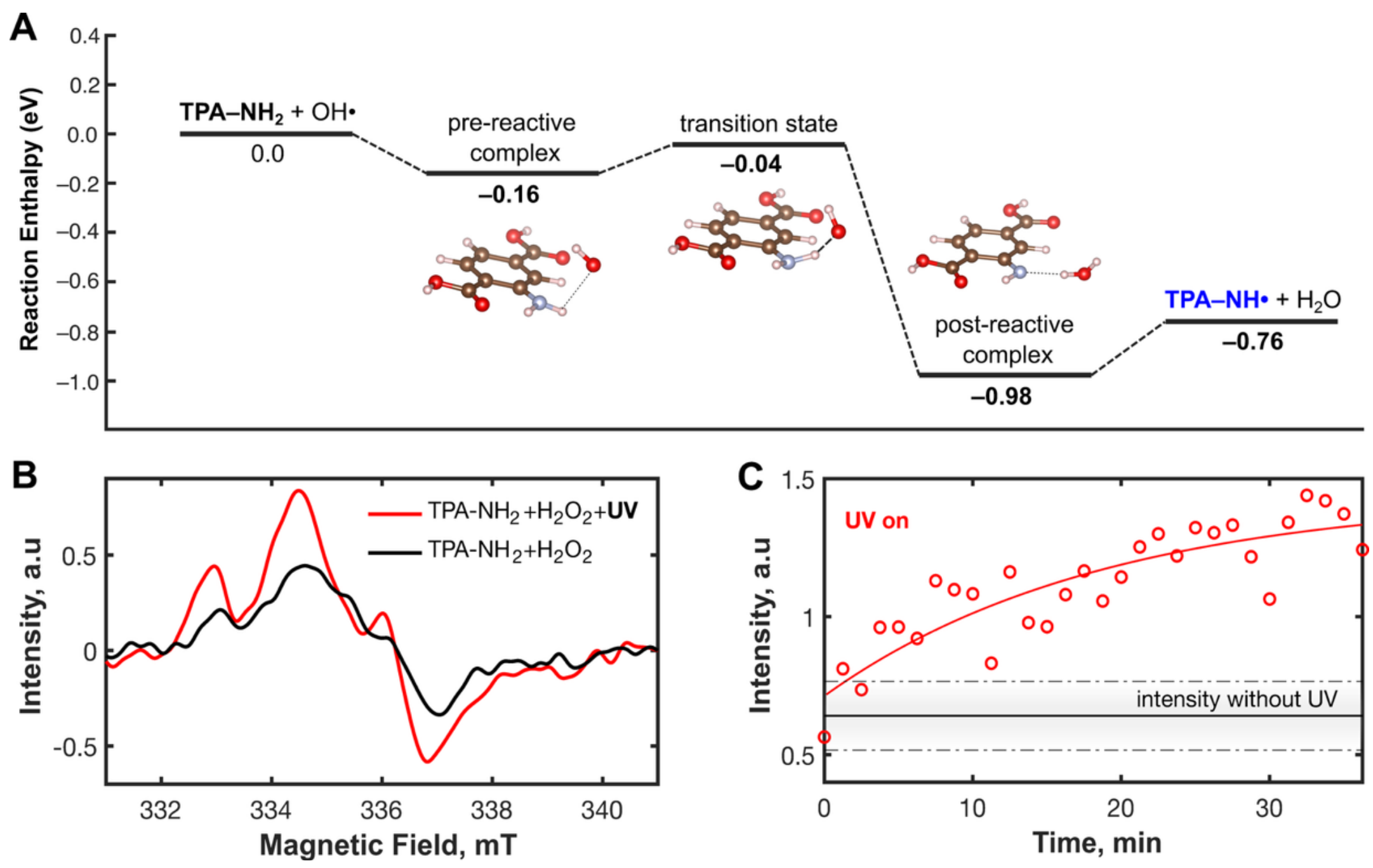

**Figure S5.** (A) Potential energy profile of H-abstraction from the $NH_2$ group of TPA-$NH_2$ upon the reaction with OH• radical. Reaction enthalpies were computed at the M06-2X/aug-cc-pVTZ level with thermal corrections at 298 K. (B) Comparison of the EPR spectrum of TPA-$NH_2$ in $H_2O_2$ (10 M) under UV irradiation (with a broadband Dymax BlueWave 50 UV lamp (280 nm to 450 nm; *red*) and without UV (*black*), showing the effect of OH• radicals generated by $H_2O_2$ photolysis. (C) Time-dependent growth of EPR signal following UV activation (red), compared to the mean signal intensity in the absence of UV (black line ± standard deviation).

**Table S2.** Relative enthalpies ($\Delta H_{298}$), Gibbs free energies ($\Delta G_{298}$), and electronic energies (with the zero-point correction; ($\Delta E_0$) of the transition states and post-reaction complexes of H-abstraction and OH-addition reactions of OH• and TPA-$NH_2$.

| | $\Delta H_{298}$ (eV) | $\Delta G_{298}$ (eV) | $\Delta E_0$ (eV) |
|---|---|---|---|
| **H-abstraction** | | | |
| Transition state | – 0.044 | 0.398 | 0.007 |
| Post-reaction complex | – 0.979 | – 0.627 | – 0.970 |
| TPA–NH• + $H_2O$ | – 0.759 | – 0.790 | – 0.760 |
| **OH-addition** | | | |
| Transition state 1 | 0.701 | 1.120 | 0.735 |
| Adduct 1 (OH, $NH_2$, COOH, COOH) | – 0.002 | 0.444 | 0.0439 |
| Transition state 2 | 0.322 | 0.741 | 0.356 |
| Adduct 2 (OH, H, $NH_2$, COOH, COOH) | – 0.667 | – 0.213 | – 0.617 |

| Transition state 3 | 0.319 | 0.756 | 0.363 |
|---|---|---|---|
| Adduct 3 | – 0.554 | – 0.095 | – 0.502 |

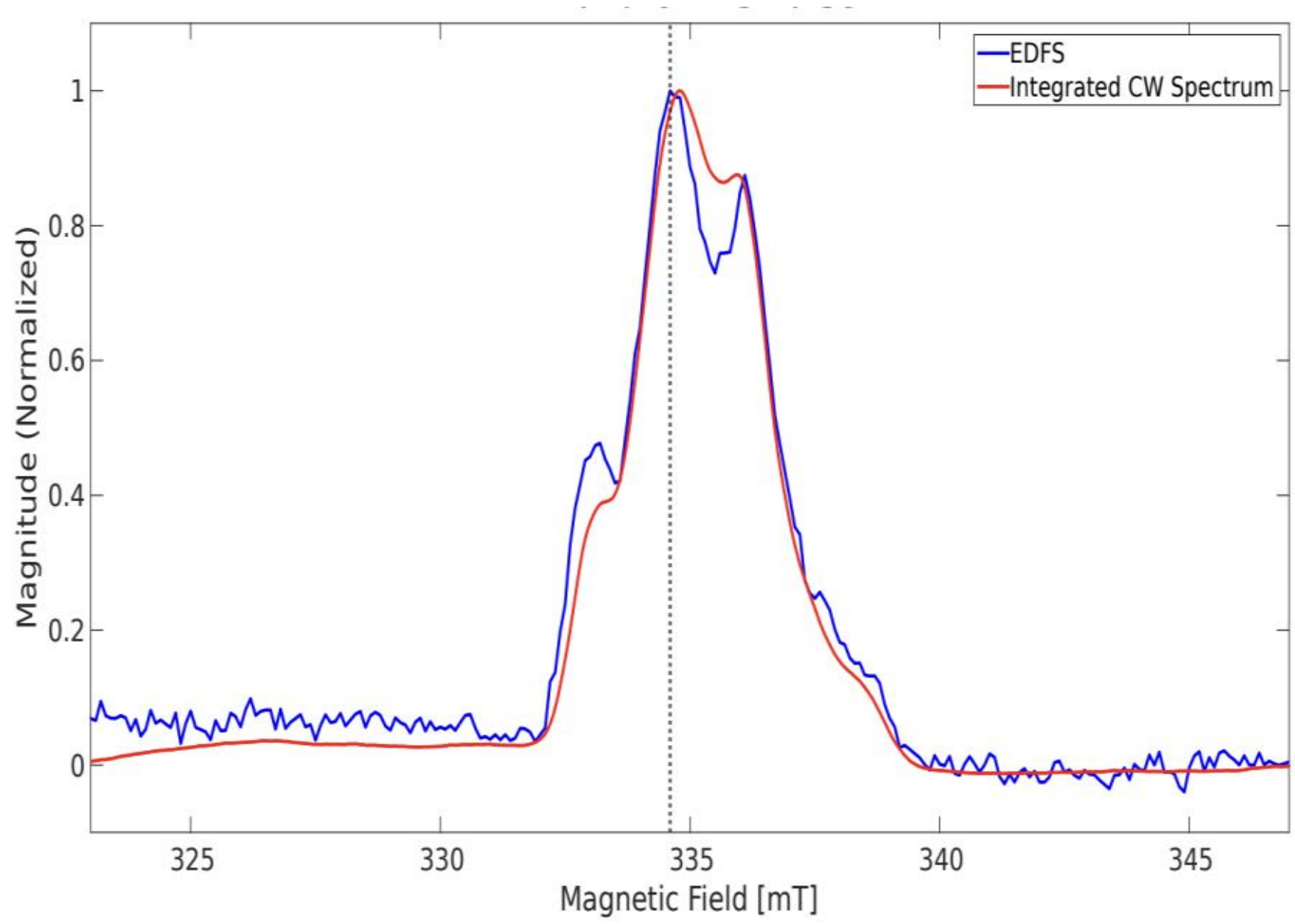

**Figure S6.** X-band EDFS spectrum of the NH• spin center in UiO-66-NH2 (blue) measured using the Hahn-echo microwave pulse sequence at a temperature $T$ = 50 K, shown in comparison with the integrated X-band CW EPR spectrum (red). A dotted line indicates the magnetic field used for all further pulsed EPR experiments.

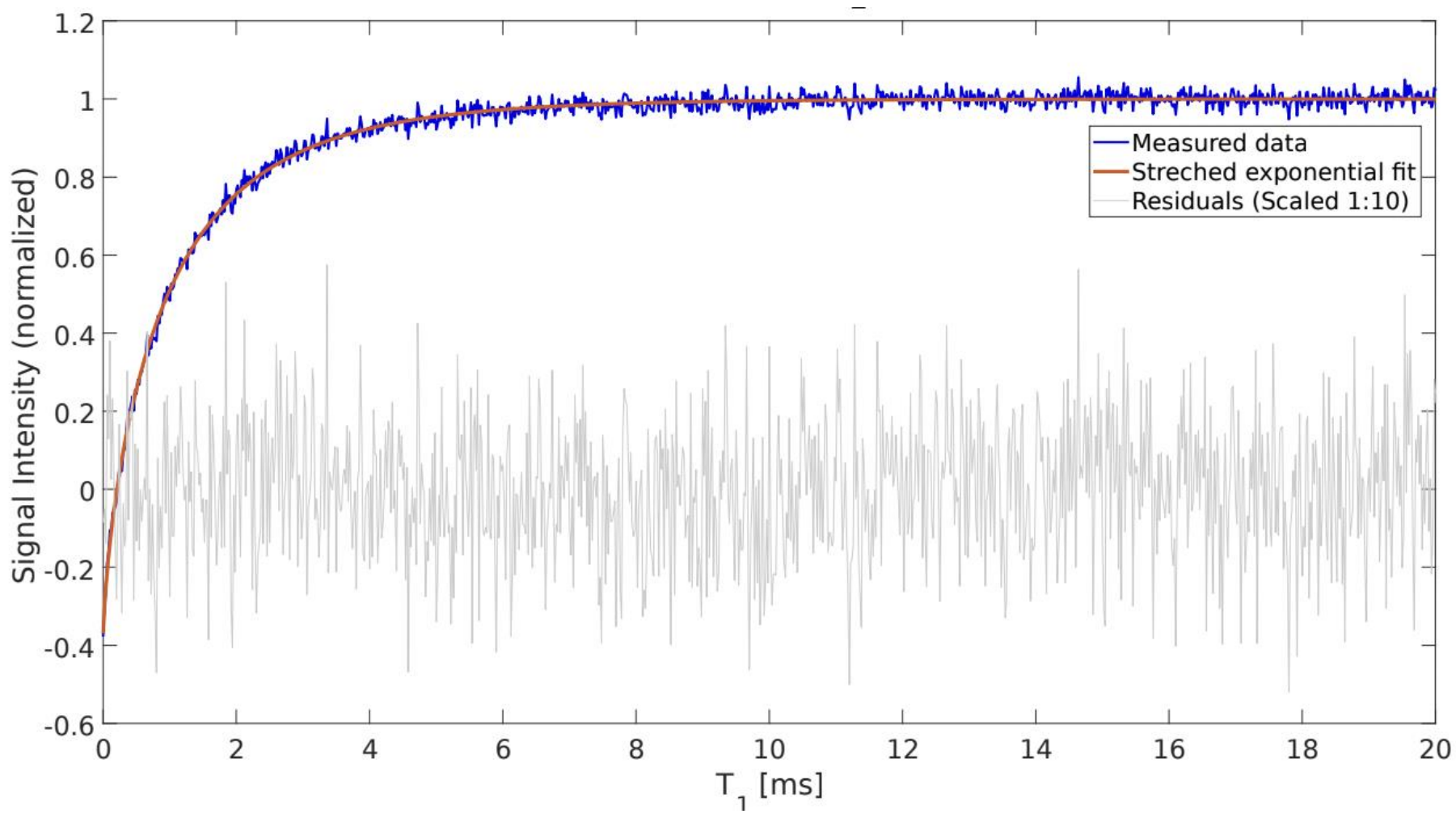

**Figure S7.** The spin-lattice relaxation time ($T_1$) was determined at 50 K using the inversion recovery pulse sequence. This relaxation time reflects how quickly the spin system returns to thermal equilibrium with its lattice environment after being excited. The data were fit using a stretched exponential model, yielding a $T_1$ of 956 µs with a stretching parameter (β) of 0.74.

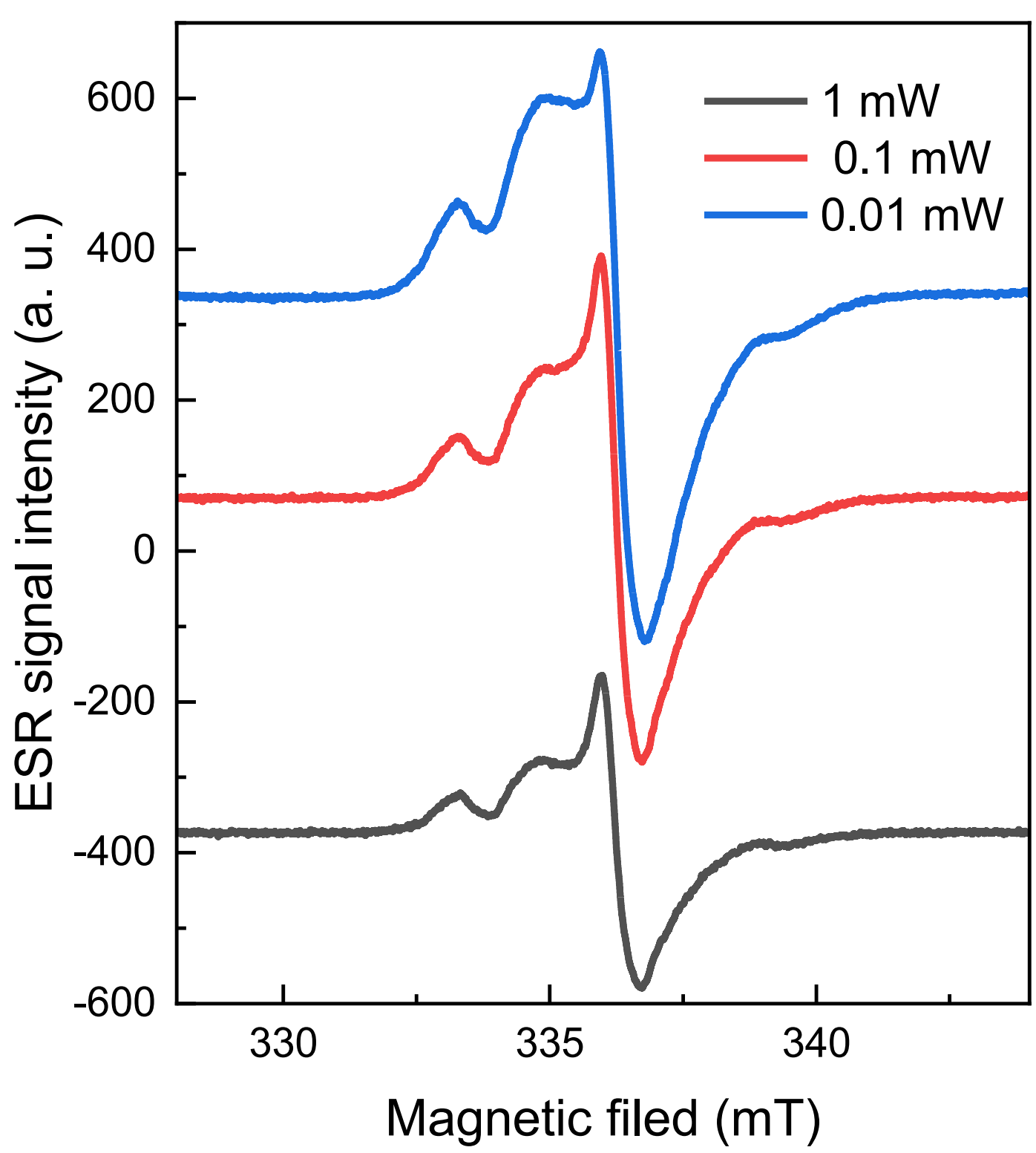

**Figure S8.** The microwave dependence of EPR signals.

## REFERENCES (SI)